\documentclass[aps,prd,preprintnumbers,showpacs]{revtex4}
\usepackage[dvips]{graphicx}

\begin{document}

%
%  Uncomment following two lines and one below for 2 column format.
%@twocolumnfalse\endcsname

\eprint{Nisho-3-2021}
\title{Detectable Electric Current induced by Dark Matter Axion in a Conductor }
\author{Aiichi Iwazaki}
\affiliation{International Economics and Politics, Nishogakusha University,\\ 
6-16 3-bantyo Chiyoda Tokyo 102-8336, Japan }   
\date{Nov. 18, 2021}
\begin{abstract}
We propose a way of detecting dark matter axion by using two slabs of conductor. 
The flat surfaces are put to meet face to face so that they are parallel to each other. 
External magnetic field $B$ parallel to the surfaces is impressed.
Radiations converted from the axion arise between two slabs.
When we tune the spacing $l$ between two surfaces such as $l=\pi/m_a$ with axion mass $m_a$,
a resonance occurs so that the radiations become strong. Furthermore,
electric current flowing on the surface of the slabs is enhanced.
We show that the electric current is large enough to be detectable at the resonance.
It reaches 
$0.7\times 10^{-9}$A$(10^{-5}\mbox{eV}/m_a)^{1/2}(B/5\mbox{T})(L/10\mbox{cm})(\sigma/3.3\times 10^7\rm eV)$, 
using $6$N copper of the 
square slab with side length $L$ and high electrical conductivity $\sigma$ at temperature $T\sim 1$K. 
The power of the Joule heating is 
$0.3\times10^{-22}\mbox{W}(B/5\mbox{T})^2(10^{-5}\mbox{eV}/m_a)^{1/2}(L/10\mbox{cm})^2(\sigma/3.3\times 10^7\rm eV)$.
When we amplify the power using LC circuit with $Q_{LC}$ factor, 
the signal to noise ratio is $4.5\times 10^{4}(Q_{LC}/10^6)(B/5\mbox{T})^2(t_{obs}/1\sec)^{1/2}\,(10^{-5}\mbox{eV}/m_a)
(L/10\mbox{cm})^2(\sigma/3.3\times 10^7\rm eV)$  
with $t_{obs}$ observational time.

\end{abstract}
\hspace*{0.3cm}
%\pacs{98.70.-f, 98.70.Dk, 14.80.Va, 11.27.+d \\
%Axion, Fast Radio Burst, Accretion disk}

\hspace*{1cm}

\maketitle

%\vskip2pc
%\section{introduction}
%%%%%%%%%%%%%%%%%%%%%%%%%%%%%%%%%%%%%%%%%%%%%%
The axion\cite{axion, Wil} is one of the most important candidates of the dark matter.
Especially, it appears that it stands on the top of the candidates because 
QCD axion gives natural solutions for strong CP problem as well as the dark matter problem. 
Other candidates of the dark matter 
e.g. WINPs ( weakly interacting massive particles )  
have not been detected in spite of extensive searches for long terms.  

The axion has recently been searched with various ways.
Most of the experiments\cite{admx, carrack, haystac, abracadabra, organ, madmax, brass, cast, sumico}  exploit the Primakoff effect. That is, 
they detect photon directly produced by the dark matter axion $a(t)$ under strong magnetic field $B$.
But, the axion photon coupling $g_{a\gamma\gamma}a(t)\sim 10^{-19}$ is so tiny that the direct photon production is
hard to be detected. So we need, for instance, very sensitive radio receiver using 
fine technology such as SQUIDs.

\begin{figure}[htp]
\centering
\includegraphics[width=0.6\hsize]{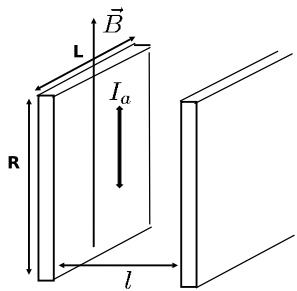}
\caption{rectangular flat conductors with size $L\times R$ and spacing $l$. Electric current $I_a$ flows
on the surface.}
\label{1}
\end{figure}

\vspace{0.1cm}
In this letter we propose a new way of detection of the dark matter axion. The detection is performed by
observing electric current flowing a conductor with high electrical conductivity. The current is induced by
the axion under strong magnetic field $\vec{B}$. We use two parallel slabs\cite{iwazaki1} Fig.\ref{1} which are put to meet face to face.
When we impress magnetic field between the slabs, the axion is converted to radiation between the slabs.
Simultaneously, electric current $I_a$ flows the surfces of the slabs
in the direction parallel to the magnetic field $\vec{B}$.
When we appropriately tune a spacing $l$ between the parallel slabs, 
a resonance occurs and the radiation is enhanced. At the resonance, the electric current
is also enhanced. We find that the electric current reaches 
$\sim 10^{-10}$A$(10^{-5}\mbox{eV}/m_a)^{1/2}(B/5\mbox{T})(L/10\mbox{cm})^2(\sigma/3.3\times 10^7\rm eV)$, using
high quality $6$N copper of the square slab with the side length $L=10$cm 
and high electrical conductivity $\sigma\simeq 3.3\times 10^7$eV at $T=1$K.
This system resembles the system used in ADMX, where conductor of cylinder form is used to obtain the resonance.
In addition to the amplification of radiation by the tuning the spacing,
we also amplify the electric current with LC circuit\cite{sikivie,kishimoto} by tuning parameters of the circuit. Consequently,
we can obtain sufficiently large power 
$\sim 10^{-17}\mbox{W}(Q_{LC}/10^6)(10^{-5}\mbox{eV}/m_a)^{1/2}(B/5\mbox{T})^2(L/10\mbox{cm})^2(\sigma/3.3\times 10^7\rm eV)$ to be detected
by the circuit with $Q_{LC}$ factor.
%\section{electric fields induced by axion in vacuum under magnetic field}
%\label{2}

\vspace{0.2cm}
Before we discuss radiation in the space between two slabs of normal conductors, 
we briefly explain the axion photon coupling.
The axion $a(\vec{x},t)$ couples with both electric $\vec{E}$ and magnetic fields $\vec{B}$ in the following,

\begin{equation}
\label{L}
L_{aEB}= g_{\gamma}\alpha \frac{a(\vec{x},t)\vec{E}\cdot\vec{B}}{f_a\pi}\equiv g_{a\gamma\gamma} a(\vec{x},t)\vec{E}\cdot \vec{B}
\end{equation}
with the decay constant $f_a$ of the axion
and the fine structure constant $\alpha\simeq 1/137$,   
where the numerical constant $g_{\gamma}$ depends on axion models; typically it is of the order of one.
The standard notation $g_{a\gamma\gamma}$ is such that $g_{a\gamma\gamma}=g_{\gamma}\alpha/f_a\pi\simeq 0.14(m_a/\rm GeV^2)$
for DFSZ model\cite{dfsz} and $g_{a\gamma\gamma}\simeq -0.39(m_a/\rm GeV^2)$ for KSVZ model\cite{ksvz}.
In other words, $g_{\gamma}\simeq 0.37$ for DFSZ and $g_{\gamma}\simeq -0.96$ for KSVZ.
%Hereafter we set $k_a=1$. 
The axion decay constant $f_a$ is related with the axion mass $m_a$ in the QCD axion; $m_af_a\simeq 6\times 10^{-6}\rm eV\times 10^{12}$GeV.
We only consider the QCD axion in this paper.

\vspace{0.1cm}
Because de Broglie wave length of the axion inside our galaxy is $10^3$ times larger than $m_a^{-1}$, we may neglect 
spatial dependence of the axion field in experiment at room; $a(\vec{x},t)=a(t)$.
We can see that
the coupling parameter $g_{a\gamma\gamma} a(t)$ in the Lagrangian eq(\ref{L}) is extremely small for the dark matter axion $a(t)$.
We note that the energy density of the dark matter axion taken by time average is given by

\begin{equation}
\rho_a= \frac{1}{2}\overline{(\dot{a}^2+(\vec{\partial} a)^2+m_a^2a^2)}\simeq \frac{m_a^2a_0^2}{2}
\end{equation}
where $a(t)=a_0\cos(t\sqrt{m_a^2+k^2} )\simeq a_0\cos(m_a t)$, because the velocity $k/m_a$ of the axion is about $10^{-3}$ in our galaxy.

The local energy density $\rho$ of dark matter in our galaxy is supposed such as $\rho\simeq 0.3\rm GeV\,\rm cm^{-3}\simeq 2.4\times 10^{-42}\rm GeV^4$.
Assuming that the density is equal to that of the dark matter axion, i.e. $\rho=\rho_a$,
we find extremely small parameter $g_{a\gamma\gamma}a\sim 10^{-19}$.
The energy density also gives the large number density of
the axions $\rho_a/m_a\sim 10^{15}\mbox{cm}^{-3}(10^{-6}\mbox{eV}/m_a)$, which give rise to their coherence. 
This allows us to treat the axions as the classical axion field $a(t)\propto \cos(m_at)$ spatially smoothly varying. 
The parameter $g_{a\gamma\gamma} a(t)$ is extremely small so that perturbative analysis is valid.

\vspace{0.2cm}
The interaction term in eq(\ref{L}) between axion and electromagnetic field slightly modifies Maxwell equations in vacuum
between two slabs,

\begin{eqnarray}
\label{modified}
\vec{\partial}\cdot\vec{E}+g_{a\gamma\gamma}\vec{\partial}\cdot(a(t)\vec{B})&=0&, \quad 
\vec{\partial}\times \Big(\vec{B}-g_{a\gamma\gamma}a(t)\vec{E}\Big)-
\partial_t\Big(\vec{E}+g_{a\gamma\gamma}a(t)\vec{B}\Big)=0,  \nonumber  \\
\vec{\partial}\cdot\vec{B}&=0&, \quad \vec{\partial}\times \vec{E}+\partial_t \vec{B}=0.
\end{eqnarray}
From the equations, we approximately obtain the electric field $\vec{E}$
generated by the axion $a(t)$  
under static and spatially homogeneous background magnetic field $\vec{B}$.
It is assumed to be of the order of $g_{a\gamma\gamma}a(t)B$,

\begin{equation}
\label{E}
\vec{E}_a(t)=-g_{a\gamma\gamma}a(t)\vec{B}.
\end{equation}
Other solutions are those of standard electromagnetic waves in vacuum.

\vspace{0.2cm}
Now, we discuss\cite{iwazaki1} radiations arising between two slabs and electric currents flowing the slabs.
In order to discuss electric and magnetic fields in the slabs composed of normal conductor,
we suppose conductor with trivial permeability $\mu=1$, 
dielectric permittivity $\epsilon=1$ and large electrical conductivity $\sigma$. 

\vspace{0.1cm}
The configurations Fig.\ref{1} of the conductors are in the following. The conductors with flat surfaces are put parallel to each other.
They occupy the regions $x \ge 0$ and $x \le -l$. Their surfaces are present at $x=0$ and $x=-l$, respectively. They
are uniform in $y$ and $z$ directions.
We impose spatially uniform static magnetic field $\vec{B}=(0,0,B)$ parallel to the surfaces of the conductors. 
We denote magnetic
field induced in the conductor as $\vec{B}(\mbox{inside})+\vec{B}$. 
The term $\vec{B}$ corresponds to the external magnetic field $\vec{B}$ because 
the component of the field $\vec{B}(\mbox{inside})+\vec{B}$ parallel to the surface ( at $x=0$ and $x=-l$ ) must be continuous at the surface, 
Although we assume two slabs infinitely extended in $y$ and $z$ directions,
results obtained below are valid as long as the wave length of radiation is much smaller than 
actual length of the conductors.

\vspace{0.1cm}
The electromagnetic fields in the conductor are described by the modified Maxwell equations,

\begin{eqnarray}
\label{1metal}
&&\vec{\partial}\cdot\vec{E}(\mbox{inside})+g_{a\gamma\gamma}\vec{\partial}\cdot(a(t)\vec{B}(\mbox{inside}))=0, \\ 
&&\vec{\partial}\times \Big(\vec{B}(\mbox{inside})-g_{a\gamma\gamma} a(t)\vec{E}(\mbox{inside})\Big)-
\partial_t\Big(\vec{E}(\mbox{inside})+g_{a\gamma\gamma} a(t)(\vec{B}(\mbox{inside})+\vec{B})\Big)=\vec{J}_e,  \nonumber  \\
&&\vec{\partial}\cdot\vec{B}(\mbox{inside})=0, \quad \vec{\partial}\times \vec{E}(\mbox{inside})+\partial_t \vec{B}(\mbox{inside})=0,
\end{eqnarray}
where we denote electric field induced in the conductors as $\vec{E}(\mbox{inside})$.

In eq(\ref{1metal}) we have included the current $\vec{J}_e=\sigma \vec{E}(\mbox{inside})$ induced by electric field $\vec{E}(\mbox{inside})$, 
but have neglected external current $\vec{J}_{ext}$ 
generating the background static magnetic field $\vec{B}$ ( $\vec{\partial}\times \vec{B}=\vec{J}_{ext}$ ). 
The current is outside of the system.
Here we have assumed Ohm's law $\vec{J}_e=\sigma \vec{E}(\mbox{inside})$.
Only the presence of the current $\vec{J}_e$ distinguishes Maxwell equations in conductor from those in vacuum. 

\vspace{0.1cm}
When we neglect axion contribution, we obtain magnetic field $\vec{B}$ inside the conductor.
Obviously, there is no electric field inside the conductor. Once
we take into account the axion contribution, the oscillating electric field $\vec{E}(\mbox{inside})$  is induced as well as
the magnetic field $\vec{B}(\mbox{inside})$ in addition to $\vec{B}$ 

Assuming the parameter $g_{a\gamma\gamma} a(t)$ small and 
noting that the fields $\vec{E}(\mbox{inside})$ and $\vec{B}(\mbox{inside})$
are of the order of  $g_{a\gamma\gamma} a(t)$,
we derive the equations,

\begin{equation}
\label{1max}
\vec{\partial}\cdot \vec{E}(\mbox{inside})=0, \quad \vec{\partial}\times \vec{B}(\mbox{inside})=\vec{J}_e+\partial_t(\vec{E}(\mbox{inside})-\vec{E}_a), \quad
\vec{\partial}\cdot \vec{B}(\mbox{inside})=0, \quad \mbox{and} \quad \vec{\partial}\times \vec{E}(\mbox{inside})+\partial_t \vec{B}(\mbox{inside})=0,
\end{equation}
where we have used the relation $\vec{\partial}\times \vec{E}_a=0$ because $\vec{\partial}\times \vec{B}=0$ inside the conductor.

Using the Ohm's law $\vec{J}_e=\sigma \vec{E}(\mbox{inside})$ in eqs(\ref{1max}), we derive 
the equation for $\vec{E}(\mbox{inside})$,

\begin{equation}
\label{19}
(\vec{\partial}^2-\partial_t^2)\vec{E}(\mbox{inside})=\sigma\partial_t\vec{E}(\mbox{inside})-\partial_t^2\vec{E}_a
\end{equation}
where we note that $\vec{E}_a\propto \cos(m_at)$. Then, we find a solution for $x>0$ ( later we discuss the solution for $x<-l$ ),

\begin{equation}
\label{21}
\vec{E}(\mbox{inside})\simeq \vec{E}_1\exp(-\frac{x}{\delta_e})\Big(\cos(\omega t-\frac{x}{\delta_e})+A\sin(\omega t-\frac{x}{\delta_e})\Big)
+\frac{1}{\sigma}\partial_t \vec{E}_a,
\end{equation} 
with arbitrary field strength $\vec{E}_1$, parameter $A$ and frequency $\omega$. The skin depth $\delta_e$ is given by $\delta_e=\sqrt{2/\sigma\omega}$.
In the derivation, we have neglected the term $\partial_t^2\vec{E}(\mbox{inside})$ in the left hand side of eq(\ref{19}), 
which is much smaller than the term $\sigma\partial_t\vec{E}(\mbox{inside})$
in the right hand side.
Namely, we have used the inequality $\sigma \gg \omega \, ( \sim m_a) $ 
because the electric conductivity $\sigma \sim 10^4$eV of copper is much larger than 
the axion mass $m_a=10^{-6}$eV $\sim 10^{-3}$eV under consideration.
We note that the parameter $m_a\delta_e=\sqrt{2m_a/\sigma}$ is much small such as $m_a\delta_e\sim 10^{-4}$ for $m_a\sim 10^{-5}\rm eV$
and $\sigma \sim 10^3$eV.

The first term of the solution represents radiation non propagating in $z$ and $y$ directions. It is only present in the surface of the conductor
due to the skin effect. It induces electric current $\sigma\vec{E}(\mbox{inside})$ on the surface.
Although $\omega$ is arbitrary at this stage, it turns out from boundary conditions at $x=0,-l$ that it is equal to $m_a$.
Hence, we put $\omega=m_a$ hereafter. Furthermore, the boundary conditions make the electric field $\vec{E}(\mbox{inside})$ be
parallel to $\vec{B}$; $\vec{E}_1=(0,0,E_1)$ just as $\vec{E}_a=-g_{a\gamma\gamma}a(t)\vec{B}$.

The second term represents
the oscillating electric field $\frac{1}{\sigma}\partial_t \vec{E}_a$ present uniformly inside the conductor. 
This term is absent when we consider conductor with finite length in $z$ direction 
because electric charges on two ends of upper and lower surfaces 
induced by the electric field screen the field $\frac{1}{\sigma}\partial_t \vec{E}_a$.
Hereafter, we neglect the term $\frac{1}{\sigma}\partial_t \vec{E}_a$.  

We also have magnetic field $\vec{B}(\mbox{inside})=(0,B_y(\mbox{inside}),0)$ 
corresponding to the electric field $\vec{E}(\mbox{inside})=(0,0,E_z(\mbox{inside}))$
because $\vec{\partial}\times \vec{E}(\mbox{inside})=-\partial_t \vec{B}(\mbox{inside})$,

\begin{eqnarray}
\label{mag}
B_y(\mbox{inside})&=&-\frac{E_1}{\delta_e m_a}\exp\big(-\frac{x}{\delta_e}\big)\Big(\sin(m_a t-\frac{x}{\delta_e})
+\cos(m_a t-\frac{x}{\delta_e})\Big) \nonumber \\
&-&\frac{AE_1}{\delta_e m_a}\exp\big(-\frac{x}{\delta_e}\big)\Big(\sin(m_a t-\frac{x}{\delta_e})
-\cos(m_a t-\frac{x}{\delta_e})\Big).
\end{eqnarray}

As a whole, we find the magnetic field $\vec{B}(\mbox{inside})+\vec{B}$ as well as the electric field $\vec{E}(\mbox{inside})$
inside the conductor ( $x>0$ ).

The strength $\vec{E}_1$ ( frequency $\omega$ and parameter $A$ ) is determined by the boundary conditions at the surfaces $x=0$
and $x=-l$.
The boundary conditions we impose
demand that the parallel components of electric and magnetic fields 
in the both sides of the boundary must be continuous. That is,  $\vec{E}(\mbox{inside})=\vec{E}_a+\vec{E}(\mbox{outside})$ and
 $\vec{B}(\mbox{inside})=\vec{B}(\mbox{outside})$, where both of the fields $\vec{E}(\mbox{outside})$ and $\vec{B}(\mbox{outside})$
represent standing waves arising in the space between two slabs. It implies that
we only consider the mode of electric field $\vec{E}(\mbox{outside})=(0,0,E_z(\mbox{outside}))$ and 
magnetic field $\vec{B}(\mbox{outside})=(0,B_y(\mbox{outside}),0)$
arising between two slabs. 
They represent the lowest TM mode non propagating in $z$ and $y$ direction.
They depends only on the coordinate $x$ and $t$.
Then, the solutions of the Maxwell equations in vacuum representing the mode are

\begin{eqnarray}
\label{st}
E_z(\mbox{outside})&=&-b_1\cos(m_ax+\delta_1)\cos(m_at)+b_2\cos(m_ax+\delta_2)\sin(m_at) \\
B_y(\mbox{outside})&=&b_1\sin(m_ax+\delta_1)\sin(m_at)+b_2\sin(m_ax+\delta_2)\cos(m_at) 
\end{eqnarray}
with constants $b_{1,2}$ and $\delta_{1,2}$. Additionally, we have
the electric field $\vec{E}_a=-g_{a\gamma\gamma}a(t)\vec{B}=
\big(0,0,-g_{a\gamma\gamma}a_0\cos(m_at) B\big)\equiv \big(0,0,E_a\cos(m_at)\big)$ induced by axion between the slabs;
$E_a\equiv-g_{a\gamma\gamma}a_0B$.

\vspace{0.1cm}

We determine the constants, $E_1$, $A$, $b_{1,2}$ and $\delta_{1,2}$  by imposing the boundary conditions,  
$E_z(\mbox{inside})=E_z(\mbox{outside})+E_a$, $B_y(\mbox{inside})=B_y(\mbox{outside})$ at $x=0$.  

\begin{eqnarray}
\label{b}
b_1\sin\delta_1&=&-\frac{E_1(1+A)}{m_a\delta_e}, \quad b_2\sin\delta_2=-\frac{E_1(1-A)}{m_a\delta_e} 
\nonumber \\
E_a-b_1\cos\delta_1&=&E_1, \quad b_2\cos\delta_2=A E_1.
\end{eqnarray}

Additionally, by imposing the boundary condition at $x=-l$, we obtain

\begin{eqnarray}
b_1\sin(\delta_1-m_al)&=&+\frac{E_1(1+A)}{m_a\delta_e} \quad 
b_2\sin(\delta_2-m_al)=+\frac{E_1(1-A)}{m_a\delta_e} \\
E_a-b_1\cos(\delta_1-m_al)&=&E_1 \quad 
b_2\cos(\delta_2-m_al)= AE_1
\end{eqnarray}

In order to obtain the relations from the boundary condition at $x=-l$, 
we note that the solutions of the electric and magnetic fields inside the conductor $x \le -l$ are given by

\begin{eqnarray}
E_z(\mbox{inside},l)&=&
+E_1\exp\Big(\frac{x+l}{\delta_e}\Big)\Big(\cos(\omega t+\frac{x+l}{\delta_e})+A\sin(\omega t+\frac{x+l}{\delta_e})\Big) \nonumber \\
B_y(\mbox{inside},l)&=&+\frac{E_1}{\delta_e m_a}\exp\big(\frac{x+l}{\delta_e}\big)\Big(\sin(m_a t+\frac{x+l}{\delta_e})
+\cos(m_a t+\frac{x+l}{\delta_e})\Big) \nonumber \\
&+&\frac{AE_1}{\delta_e m_a}\exp\big(\frac{x+l}{\delta_e}\big)\Big(\sin(m_a t+\frac{x+l}{\delta_e})
-\cos(m_a t+\frac{x+l}{\delta_e})\Big) \nonumber \\
&\mbox{for}& \quad  x \le -l 
\end{eqnarray}

\vspace{0.1cm}
Therefore, 
we find that the coefficients, $E_1$ and $AE_1$ as well as $b_{1,2}$ are given  
such that 

\begin{eqnarray}
\label{coef}
E_1&=& E_a\frac{(m_a\delta_e\tan\delta)^2-m_a\delta_e\tan\delta}{1+(m_a\delta_e\tan\delta-1)^2}
\quad E_1A=-E_a\frac{m_a\delta_e\tan\delta}{1+(m_a\delta_e\tan\delta-1)^2} \nonumber \\ 
\quad b_1&=&\frac{E_a}{\cos\delta}\,\frac{2-m_a\delta_e\tan\delta}{1+(m_a\delta_e\tan\delta-1)^2}, 
\quad b_2= -\frac{E_a}{\cos\delta}\,\frac{m_a\delta_e\tan\delta}{1+(m_a\delta_e\tan\delta-1)^2},
\end{eqnarray}
and the phases $\delta_{1,2}=\delta$ satisfying the identical equation,

\begin{equation}
\tan\delta=\frac{\sin(m_al)}{1+\cos(m_al)}.
\end{equation}

Therefore, the coefficients behave in the limit of large electrical conductivity $\sigma$ 
( small skin depth $m_a\delta_e \ll 1$ ) such that

\begin{equation}
E_1= E_a\times O(m_a\delta_e), \quad A=-1+O(m_a\delta_e), \quad b_1=\frac{E_a}{\cos\delta_1}\Big(1+O(m_a\delta_e)\Big), \quad 
b_2=-\frac{E_a}{\cos\delta_1}O(m_a\delta_e).
\end{equation}

We find that the standing waves in eq(\ref{st}) of the electromagnetic fields in the space between the two slabs are
given by in the limit $m_a\delta_e \to 0$ ( $\delta_e=\sqrt{2/m_a\sigma}$ ) for $\sigma \to \infty$ ,
\begin{eqnarray}
\label{field}
E_z(\mbox{outside})&=&-\frac{E_a}{\cos\delta}\cos(m_ax+\delta)\cos(m_at) \\
B_y(\mbox{outside})&=&\frac{E_a}{\cos\delta}\sin(m_ax+\delta)\sin(m_at).
\end{eqnarray} 

In addition to the standing waves, we have electric field $E_z(\mbox{inside})$ in the conductor, 

\begin{equation}
\vec{E}(\mbox{inside})=(0,0,E_z), \quad E_z=-\frac{E_am_a\delta_e\tan\delta}{\sqrt{2}}\exp(-\frac{x}{\delta_e})\sin(m_at-\frac{x}{\delta_e}+\frac{\pi}{4})
\end{equation}
which generates electric current $J=\sigma E_z(\mbox{inside})$ in the surface of the conductor.
These are solutions in the limit of perfect conductor.

Because $\cos\delta_1$ is of the order of $1$, 
we find that the strengths of the standing waves in the space between the two slabs are identical to the electric field $E_a$
generated by the axion in vacuum. It is similar to the case in resonant cavity, as has been pointed out in our previous paper\cite{iwazaki1}.

\vspace{0.1cm}
We note that in the limit of perfect conductor with $\sigma\to \infty$,
the electric field $E_a+E_z(\mbox{outside})$ present between the slabs vanishes at the boundaries at $x=0$ and $x=-l$.
Similarly we find that the electric field $E(\mbox{inside})\sim E_a\times m_a\delta_e$
inside the conductors vanishes in the limit of the perfect conductor.
The magnetic field $B_y(\mbox{inside})\propto \exp(-x/\delta_e)$ also vanishes.
On the other hand, the magnetic field $B_y(\mbox{outside})$
does not vanish at the boundaries, $B_y(\mbox{outside})=E_a\tan\delta_1 \sin(m_at)$ at $x=0$ and $x=-l$.
The discontinuity takes place because of the presence of surface current.
For instance, the surface current at $x=0$ is given by

\begin{equation}
J_a=\sigma E(\mbox{inside}) \to -E_a\tan\delta\sin(m_at)\delta(x) \quad \mbox{for} \quad \sigma \to \infty .
\end{equation}
This surface current generates the discontinuity 
$\Delta B\equiv B_y(\mbox{outside})-B_y(\mbox{inside})=E_a\tan\delta \sin(m_at)$ at the surface. 

\vspace{0.2cm}
We would like to make a comment on the results\cite{iwazaki} in our previous papers, in which electromagnetic radiations from a cylindrical conductor 
converted from the dark matter axion have been discussed. In the papers we have naively derived electric current flowing its surface 
induced by the axion without appropriately imposing the boundary conditions; the boundary conditions are 
carefully treated in the present paper as well as 
our paper\cite{iwazaki1}.  As a result, 
the radiations from the conductor have been overestimated. It has already been indicated in the paper\cite{iwazaki1}. 
Their electric fields do not vanish at the surface 
even in the limit of perfect conductor. The careful treatment\cite{iwazaki2} of the boundary conditions lead to
physically reasonable view: The radiations $\vec{E}_a$ arising in vacuum from the axion under magnetic field collides the conductor and
reflected radiations are just ones emitted from the conductor. The reflected radiations are never enhanced
compared with the incident radiations $\vec{E}_a$.
The identical conclusion has also been obtained in the paper\cite{kishimoto} by Kishimoto and Nakayama.

\vspace{0.2cm}

Now we show that a resonance occurs when we adjust spacing $l$ just as $l=\pi/m_a$ in the case of finite $\sigma$.
( The results shown below hold even for $l=\pi(2n+1)/m_a$ with non negative integer $n$. ) 
The resonance implies that electric and magnetic field of radiation is enhanced compared with the radiation in eq(\ref{field}). 
With the choice of $l=\pi/m_a$, we obtain $\delta=\pi/2$ so that $ A=0, \quad E_1=E_a, \quad b_{1,2}=-\frac{E_a}{m_a\delta_e}$.
Therefore, the radiation fields $E_z(\mbox{outside})$ and $B_y(\mbox{outside})$ are enhanced
with the enhancement factor $1/m_a\delta_e$,
\begin{eqnarray}
E_z(\mbox{outside})&=&\frac{\sqrt{2}E_a}{m_a\delta_e}\sin(m_ax)\sin(m_at-\frac{\pi}{4}) \\
B_y(\mbox{outside})&=&-\frac{\sqrt{2}E_a}{m_a\delta_e}\cos(m_ax)\sin(m_at+\frac{\pi}{4}).
\end{eqnarray}

We also have electric field inside the conductor $x>0$

\begin{equation}
E_z(\mbox{inside})=E_a\exp(-\frac{x}{\delta_e})\cos(m_at-\frac{x}{\delta_e}),
\end{equation}
which induces large oscillating electric current $\sigma E_a\exp(-\frac{x}{\delta_e})\cos(m_at-\frac{x}{\delta_e})$
in the surface.

\vspace{0.1cm}
We estimate the electric current $I_a(m_at)=L\int_0^{\infty} dx \sigma E_a \exp(-x/\delta_e)\cos(m_at-x/\delta_e)=
L\sigma \delta_e E_a\sin (m_at+\pi/4)/\sqrt{2}\equiv I_a\sin(m_at+\pi/4)$ with the length $L$ 
of the slab in $y$ direction,

\begin{equation}
I_a=\frac{\sigma \delta_e E_aL}{\sqrt{2}}=\frac{\sigma \delta_e g_{a\gamma\gamma}a_0BL}{\sqrt{2}}
=0.7\times 10^{-9}\mbox{A}\Big(\frac{10^{-5}\rm eV}{m_a}\Big)^{1/2}
\Big(\frac{B}{5 \rm T}\Big)\Big(\frac{L}{10\rm cm}\Big)\Big(\frac{\sigma}{3.3\times 10^7\rm eV}\Big)
\end{equation}
where we have adopted high quality $6$N copper with electrical conductivity $\sigma \simeq 3.3\times 10^7$eV at low temperature of the order of $1$K.
We have used the parameters of the dark matter axion energy density $\rho_a=0.3\mbox{GeV}\,\mbox{cm}^{-3}=m_a^2a_0^2/2$
and the axion model $g_{\gamma}=1$ ( $g_{a\gamma\gamma}=g_{\gamma}\alpha/\pi f_a$ ). These values are used 
throughout the paper.

\vspace{0.1cm}
The power of Joule heating by the electric current $I_a$ is given by 

\begin{eqnarray}
P_a&=&LR\sigma E_a^2\int_0^{\infty}dx \exp(-\frac{2x}{\delta_e})\,
\overline{\cos^2(m_at-x/\delta_e)}=\frac{LR\delta_e\sigma E_a^2}{4} \nonumber \\
&\simeq& 0.3\times 10^{-22}\mbox{W}
\Big(\frac{10^{-5}\rm eV}{m_a}\Big)^{1/2}\Big(\frac{B}{5 \rm T}\Big)^2\Big(\frac{\sigma}{3.3\times 10^7\rm eV}\Big)
\Big(\frac{L}{10\rm cm}\Big)\Big(\frac{R}{10\rm cm}\Big)
\end{eqnarray}  
where $R$ denotes a side length of the slab ( the surface area is $R\times L$ ).
$LR\delta_e$ is the volume of the conductor in which the electric current flows. 
$\overline{O}$ is time average of $O$. 
In order to detect the electric current, we amplify the power $P_a$ using LC circuit\cite{sikivie,kishimoto}.

Because thermal noise of the system with temperature $T=1$K
is given by

\begin{equation}
P_T=\frac{T\delta \omega}{2\pi}\simeq 3.3\times 10^{-20}\mbox{W}
\Big(\frac{T}{1\rm K}\Big)\Big(\frac{\delta \omega}{10^{-6}m_a}\Big)\Big(\frac{m_a}{10^{-5}\rm eV}\Big),
\end{equation} 
the signal to noise ratio is given by

\begin{eqnarray}
\frac{S}{N}&=&\frac{P_a}{P_T}Q_{LC}\sqrt{\frac{\delta \omega t_{obs}}{2\pi}} \nonumber \\
&\simeq& 4.5\times 10^{-2} Q_{LC}\sqrt{\frac{t_{obs}}{1\rm sec}}
\Big(\frac{10^{-5}\rm eV}{m_a}\Big)\Big(\frac{B}{5 \rm T}\Big)^2\Big(\frac{\sigma}{3.3\times 10^7\rm eV}\Big)
\Big(\frac{L}{10\rm cm}\Big)\Big(\frac{R}{10\rm cm}\Big)\Big(\frac{1\rm K}{T}\Big)\Big(\frac{10^{-6}m_a}{\delta\omega}\Big)^{1/2} 
\end{eqnarray}
where $Q_{LC}$ denotes $Q$ factor in LC circuit which is used to amplify the power $P_a$.
$t_{obs}$ is the observation time.
$Q_{LC}$ is given such that $Q_{LC}=\frac{1}{R_{LC}}\sqrt{L_i/C}$ with inductance $L_i$ of coil, capacitance $C$ of condenser 
and resistance $R_{LC}$ of the circuit. Especially, $R_{LC}=R/(\sigma L\delta_e)$. Thus, we find 

\begin{equation}
Q_{LC}=\sqrt{\frac{2\sigma}{m_a}}\times \sqrt{\frac{L_i}{C}}\Big(\frac{L}{R}\Big)\simeq 2.5\times 10^6\sqrt{\frac{10^{-5}\mbox{eV}}{m_a}}
\sqrt{\frac{L_i}{C}}\Big(\frac{L}{R}\Big) .
\end{equation}

Therefore, we obtain sufficiently large $S/N$ ratio to detect the electric current induced by the dark matter axion,
even if the observational time $t_{obs}$ is $1$ second.  

\vspace{0.1cm}
We would like to point out that $Q$ value associated with the absorption by the flat conductors is 
derived using the definition $Q=m_a U/2P_a$ with the energy $U$ of radiations in the space between two slabs,

\begin{equation}
U\equiv \frac{1}{2}\int (\overline{E_z^2(\mbox{outside})}+\overline{B_y^2(\mbox{ouside})}) dV=\frac{LRE_a^2l}{2(m_a\delta_e)^2}
\end{equation}  
with $l=\pi/m_a$.
We find that 

\begin{equation}
Q=\frac{m_al}{\delta_e\sigma (m_a\delta_e)^2}=\frac{\pi}{4}\sqrt{\frac{2\sigma}{m_a}}\simeq 2\times10^6 \sqrt{\frac{10^{-5}\mbox{eV}}{m_a}}
\end{equation}
The formula is almost identical to the $Q_{LC}$ factor in the LC circuit.

\vspace{0.1cm}
In order to obtain the high value of the electric current $I_a$, we need to exactly tune the spacing $l=\pi/m_a$ between two slabs. 
But, such an exact tuning would be not possible because $\delta_e=\sqrt{2/m_a\sigma}\simeq 1.6\times 10^{-6}\mbox{cm}\sqrt{10^{-5}\mbox{eV}/m_a}$.
In actual experiment we tune it approximately. Thus,
we need to see electric current and its power in the approximate tuning.

\vspace{0.1cm}
We note that in general, electric current $I_a(l)$ flowing the slab is given by 

\begin{eqnarray}
I_a(l)&=&L\sigma E_1\int_0^{\infty}dx\exp(-\frac{x}{\delta_e})\Big(\cos(m_at-x/\delta_e)+A\sin(m_a t-x/\delta_e)\Big) \nonumber \\
&=&\frac{L\sigma \delta_e E_1}{\sqrt{2}}\Big(\sin(m_at+\frac{\pi}{4})+A\sin(m_at-\frac{\pi}{4})\Big), 
\end{eqnarray}

and the power of Joule heating $P_a(l)$ is

\begin{eqnarray}
P_a(l)&=&LR\sigma E_1^2\int_0^{\infty}dx \exp(-\frac{2x}{\delta_e})\overline{\Big(\cos(m_at-x/\delta_e)+A\sin(m_a t-x/\delta_e)\Big)^2}
\nonumber \\
&=&\frac{LR\delta_e\sigma (E_1^2+(E_1A)^2)}{4},
\end{eqnarray}
where the parameters $E_1$ and $A$ depend on the spacing $l$ between two slabs.

\vspace{0.1cm}
When we take the spacing
$l=\pi/m_a+\delta l$ with $\delta l \ll \pi/m_a$ ( then, $\tan\delta \simeq -2/(m_a\delta l)$ ), the parameters $E_1$ and $E_1A$ are found such as 

\begin{equation}
E_1=-\frac{E_a(y+1)}{2y^2+2y+1}, \quad E_1A=\frac{E_a y}{2y^2+2y+1} \quad \mbox{with} \quad y\equiv \delta l/2\delta_e. 
%\quad b_1=-\frac{E_a}{m_a\delta_e}\frac{2y+1}{2y^2+2y+1},
%\quad b_2=-\frac{E_a}{m_a\delta_e}\frac{1}{2y^2+2y+1}
\end{equation}
The tuning $y \ll 1$ would be very difficult because 
$\delta_e\simeq 1.6\times 10^{-6}\mbox{cm}\sqrt{10^{-5}\mbox{eV}/m_a}$.
But even if we tune $y$ such as $y\sim10^2$, i.e. $\delta l \sim 10^{-4}$cm, 
we obtain electric current $I_a(l=\pi/m_a+\delta l)=-I_a(\delta l)\cos(m_at)$,

\begin{equation}
I_a(\delta l)=\frac{\sigma \delta_e E_aL}{2y}\simeq 1.4\times10^{-11}\mbox{A}
\Big(\frac{10^{-4}\rm cm}{\delta l}\Big)\Big(\frac{10^{-5}\rm eV}{m_a}\Big)
\Big(\frac{B}{5 \rm T}\Big)\Big(\frac{L}{10\rm cm}\Big)\Big(\frac{\sigma}{3.3\times 10^7\rm eV}\Big),
\end{equation}
and the corresponding power $P_a(\delta l) \equiv P_a(l=\pi/m_a+\delta l)$,

\begin{equation}
P_a(\delta l)=\frac{LR\delta_e \sigma E_a^2}{8y^2} \simeq 1.5\times 10^{-26}\mbox{W}
\Big(\frac{10^{-4}\rm cm}{\delta l}\Big)^2\Big(\frac{10^{-5}\rm eV}{m_a}\Big)^{3/2}
\Big(\frac{B}{5 \rm T}\Big)^2\Big(\frac{\sigma}{3.3\times 10^7\rm eV}\Big).
\Big(\frac{L}{10\rm cm}\Big)\Big(\frac{R}{10\rm cm}\Big) 
\end{equation}

Then, we find the signal to noise ratio $S(\delta l)/N$

\begin{eqnarray}
\label{SN}
&&\frac{S(\delta l)}{N}=\frac{P_a(\delta l)}{P_T}Q_{LC}\sqrt{\frac{\delta \omega t_{obs}}{2\pi}} \nonumber \\
&\simeq& 2.3\times 10^{-5}Q_{LC}\sqrt{\frac{t_{obs}}{1\rm sec}}
\Big(\frac{10^{-4}\rm cm}{\delta l}\Big)^2\Big(\frac{10^{-5}\rm eV}{m_a}\Big)^2
\Big(\frac{B}{5 \rm T}\Big)^2\Big(\frac{\sigma}{3.3\times 10^7\rm eV}\Big)
\Big(\frac{L}{10\rm cm}\Big)\Big(\frac{R}{10\rm cm}\Big)\Big(\frac{1\rm K}{T}\Big)\Big(\frac{10^{-6}m_a}{\delta \omega}\Big)^{1/2} 
\end{eqnarray}
where $Q_{LC}=\frac{1}{R_{LC}}\sqrt{L_i/C}\sim 10^6\sqrt{L_i/C}\sqrt{10^{-5}\mbox{eV}/m_a}$ with the use of $L=10$cm and $R=10$cm.

%Therefore, we can determine axion mass $m_a$ by 
%detecting the electric current $I_a$ with tuning the spacing $l=\pi/m_a$ between two slabs. 

\vspace{0.1cm}
In the actual experiment with the use of $Q_{LC}=10^6$, we can search the spacing $\Delta l\sim 6$cm within $1$ day,
if we tune the spacing with the speed $\partial_t\delta l \sim10^{-4}\rm cm/s$ ( $t_{obs}=1$sec for $S/N>1$ ), 
It covers the range of the QCD axion mass $m_a\sim 10^{-5}\mbox{eV}$ because $\pi/m_a\sim 6$cm. 
On the other hand, in order to cover the range of the axion mass $\sim 10^{-4}$eV ( $\pi/m_a\sim 6\times 10^{-1}$cm ),
we can search it for $10$ days if we tune the spacing with the speed $\partial_t\delta l \sim10^{-4} \rm cm/10^2s$ ( $t_{obs}=10^2$sec for $S/N>1$ ).
But, to cover the range of $m_a=10^{-3}$eV, we need $t_{obs}=10^6$sec for $S/N>1$ so that it takes more than $10$ years to 
search the spacing $\Delta l\sim 0.06$cm.
However, if we can tune the spacing $\delta l=10^{-5}$cm, by taking $\partial_t\delta l \sim10^{-5}\rm cm/10^2s$
( $t_{obs}=10^2$sec  for $S/N>1$ ), we can search the spacing $\Delta l\sim 0.06$cm within
$10$ day even for $m_a=10^{-3}$eV.

On the other hand, when we search the mass $\sim 3\times10^{-6}$eV ( $\pi/m_a\sim 20$cm ), we need larger slab such as $L=R=40$cm
so as for our approximation of infinitely extending conductor to hold. Then, the power of Joule heating or $S/N$ ratio is
much bigger so that the detection is much easier because we can take $\delta l=10^{-3}$cm and $t_{obs}=1$sec, as we can see from eq(\ref{SN}).
In this way, it turns out that the new way of the axion detection is very effective
for the QCD axion search.

\vspace{0.2cm}
The author
expresses thanks to Yasuhiro Kishimoto and Kazunori Nakayama for useful comments and discussions.
He also expresses thanks to Ngyuen, Le Hoang for useful suggestions.
This work is supported in part by Grant-in-Aid for Scientific Research ( KAKENHI ), No.19K03832.

%\vspace*{1cm}
% of KEK for their useful discussion and comments.

%%%%%%%%%%%%%%%%%%%%%%%

\end{document}